\documentclass[twocolumn,showpacs,preprintnumbers,amsmath,amssymb]{revtex4}

\usepackage{graphicx}
\usepackage{dcolumn}
\usepackage{bm}
\begin{document}

\preprint{SOC-DF-v2}

\title{Analysis of self-organized criticality in Ehrenfest's dog-flea model}

\author{Burhan Bakar$^1$}
 \email{burhan.bakar@ege.edu.tr}
\author{Ugur Tirnakli$^{1,2}$}
 \email{ugur.tirnakli@ege.edu.tr}
\affiliation{
 $^1$Department of Physics, Faculty of Science, Ege University, 35100 Izmir, Turkey\\
$^2$Division of Statistical Mechanics and Complexity, 
Institute of Theoretical and Applied Physics (ITAP) Kaygiseki Mevkii, 
48740 Turunc, Mugla, Turkey
}

\date{\today}

\begin{abstract}
The self-organized criticality in Ehrenfest's historical dog-flea model is analyzed by simulating the 
underlying stochastic process. The fluctuations around the thermal equilibrium in the model are treated 
as avalanches. We show that the distributions for the fluctuation length differences at subsequent time 
steps are in the shape of a $q$-Gaussian (the distribution which is obtained naturally in the context 
of nonextensive statistical mechanics) if one avoids the finite size effects by increasing the system size. 
We provide a clear numerical evidence that the relation between the exponent $\tau$ of avalanche size 
distribution obtained by maximum likelihood estimation and the $q$ value of appropriate $q$-Gaussian 
obeys the analytical result recently introduced by Caruso \emph{et al.} 
[Phys. Rev. E \textbf{75}, 055101(R) (2007)]. This rescues the $q$ parameter to remain as a fitting 
parameter and allows us to determine its value \emph{a priori} from one of the well known exponents 
of such dynamical systems.        
\end{abstract}


\maketitle

%%%%%%%%%%%%%%%%%%%%%%%%%%%%%%%%%%%%%%%%
%\section{\label{sec:Int}Introduction}
%%%%%%%%%%%%%%%%%%%%%%%%%%%%%%%%%%%%%%%%%
{\it Introduction:}$\,\,$ The term self-organized criticality (SOC) was first introduced by Bak, Tang, and Wiesenfeld (BTW) in 1987 \cite{Bak-PRL59}. In their well known paper, the so-called BTW sandpile model was used to demonstrate that the dynamics which gives rise to the power-law correlations seen in the non-equilibrium steady states must not involve any fine-tuning of parameters. Namely, systems under their natural evolution are driven at a very slow rate until one of their elements reaches a threshold, \emph{i.e.}, statistically stationary state, and this triggers a burst of activity (avalanche) which occurs on a very short time scale. When the avalanche is over, the system evolves again according to the slow drive until a next avalanche is 
triggered. The activity of the system in this way consists of a series of avalanches. There are many systems where the SOC paradigm has been applied, \emph{e.g.} earthquakes, noise with $1/f$ power spectrum, brain activity, river networks, biological evolution of interacting species, traffic jams \emph{etc.} \cite{Jensen-BakBook}. 

Following the BTW sandpile model a great variety of models from the deterministic and stochastic to the dissipative and conservative have been introduced which exhibit the phenomenon of SOC (for an overview, see \cite{Dhar} and references therein). In 1996, a random neighbor version of the original BTW sandpile model was presented by Flyvbjerg \cite{Flyvbjerg-PRL}. In this work, it was emphasized that a self-organized critical system is a driven, dissipative system consisting of a medium (sandpile) which has disturbance propagating through it, causing a modification of the medium, such that eventually the medium 
is in a critical state, and the medium is modified no more. Moreover, it was shown by way of random neighbor sandpile model that a dynamical system with only two degrees of freedom can be self-organized critical and as it is the case in fluctuation phenomena, the dynamics is described by a master equation which can be partially solved analytically. 

Soon after Flyvbjerg's work Nagler \emph{et al.} studied the conservative variant of random neighbor sandpile 
model which is neither extended nor dissipative with regard to the amount of sand in the system but still 
shows SOC with nontrivial exponents \cite{Nagler-PRE, Nagler-JSP}. This kind of analysis is not restricted 
to nonspatial systems and available also for spatial systems like one-dimensional cellular 
automata \cite{Nagler3}. 
The dynamics of the model described by Nagler \emph{et al.} is given on a Fokker-Planck equation by 
introducing appropriate scaling variables. The avalanche size distribution which is readily obtained by 
solving the Fokker-Planck equation at an absorbing boundary exhibits a power-law regime followed by an 
exponential tail. Their model is an adaptation of the famous dog-flea model introduced by Ehrenfest in 
1907 \cite{Ehrenfest}. This model can be considered as a zero-dimensional nonspatial prototype SOC model 
and its dynamics is different from most of the standard SOC models which are $N$-dimensional spatial 
systems.

The dog-flea model is a simple but typical example of generation-recombination Markov chain \cite{Ehrenfest2} 
describing the process of approaching an equilibrium state in a large set of uncoupled two state systems 
together with fluctuations (avalanches) around this state. For an even number of states, the transition 
probability of fluctuations of the discrete time version was calculated by Kac \cite{Kac1} 
(see also \cite{Kac2}). An identification of the model as a random walk on a Bethe lattice is studied in 
Ref.~\cite{Hughes-Yellin-Monthus}. Furthermore, it has recently been shown that the dog-flea model, 
formulated as a continuous time Markov chain, is a representation of a spin in a magnetic field \cite{LL_PLA}. 
Such a representation is used to estimate the blocking temperature in molecular nano-magnets \cite{BBLL1}.

In this work, we will be analyzing the SOC in the dog-flea model through simulation of the underlying 
stochastic process that describes the natural evolution of the model. The analysis method that we use 
has recently been presented by Caruso \emph{et al.} to interpret the SOC in the limited number of 
earthquakes (up to 689 000) taken from World and Northern California catalogs for the periods 2001-2006 
and 1966-2006, respectively \cite{CarusoPRE75}. Using the same line of thought, it is our aim to analyze 
the SOC feature of the dog-flea model through the time series of the fluctuation length. The simplicity 
of the dynamics of the dog-flea model enables us to obtain a large number of fluctuations for different 
system sizes in a reasonable computing time (\emph{i.e.}, we consider up to $2\times10^{9}$ fluctuations). 
Thus, the obtained critical exponents for the model are very precise as it will be discussed in 
coming sections. 
This analysis enables us to accomplish our main task, which is to provide the first rigorous numerical 
example where the relationship, proposed by Caruso \emph{et al.}, between the exponent $\tau$ of avalanche size 
distribution and the $q$ value of appropriate $q$-Gaussian (the distribution which is obtained naturally 
in the context of nonextensive statistical mechanics) \cite{Tsallis88-NonExEnt}. This will be very 
appealing also from nonextensive statistical mechanics point of view since this treatment makes the 
$q$ parameter to be determined \emph{a priori}, which is a situation achieved rarely up to now.

%%%%%%%%%%%%%%%%%%%%%%%%%%%%%%%%%%%%%
%\section{\label{sec:Model} The model and numerical procedure}
%%%%%%%%%%%%%%%%%%%%%%%%%%%%%%%%%%%%%
{\it The model and numerical procedure:}$\,\,$ The dynamics of the dog-flea model has simple rules. The model 
has $N$ dynamical sites represented by the total number of fleas shared by two dogs (dog $A$ and dog $B$). 
Suppose that there are $N_{A}$ fleas on dog $A$ and $N_{B}$ fleas on dog $B$ leading to a population of fleas 
$N=N_{A}+N_{B}$. 
For convenience, $N$ is assumed to be even. In every time step, a randomly chosen flea jumps from one dog to the other. Thus, we have $N_{A}\rightarrow N_{A}\pm1$ and $N_{B}\rightarrow N_{B}\mp1$. The procedure is repeated for an arbitrary number of times. In long time run, the mean number of fleas on both dog $A$ and dog $B$ converges to the equilibrium value, $\langle N_{A}\rangle =\langle N_{B}\rangle = N/2$ with the fluctuations around it. A single fluctuation is described as a process that starts once the number of fleas on one of the dogs becomes larger (or smaller) than the equilibrium value $N/2$ and stops when it gets back to it for the first time. Thus, the end of one 
fluctuation specifies the start of the subsequent one. The length ($\lambda$) of a fluctuation is determined by the number of time steps elapsed until the fluctuation ends.    

It is straightforward to obtain the master equation of the process that describes the time evolution of the probability 
to find a specified number of fleas on one of the dogs. Assuming that after $t$ steps there are $N_{A}(t)=\ell$ fleas 
on dog $A$, at the subsequent time step there are only two possibilities, $\ell\rightarrow\ell+1$ or $\ell\rightarrow\ell-1$ with the 
transition probabilities $W(\ell+1|\ell)=(N-\ell)/N$ and $W(\ell-1|\ell)=\ell/N$, respectively. Then, the time evolution of the probability $P(\ell,\,t)$ to find $\ell$ fleas on dog $A$ at time $t$ obeys the following master equation,
\begin{equation}\label{eq:master}          
P(\ell,\,t+1)=\frac{\ell+1}{N}P(\ell+1,\,t)+\frac{N-\ell+1}{N}P(\ell-1,\,t).
\end{equation}
Introducing appropriate scaling variables Eq.~(\ref{eq:master}) can be written in the form of a Fokker-Planck equation by which the fluctuation distribution is reviewed analytically \cite{Nagler-PRE}. 

\begin{figure*}[t]
\begin{center}
\includegraphics[width=8cm]{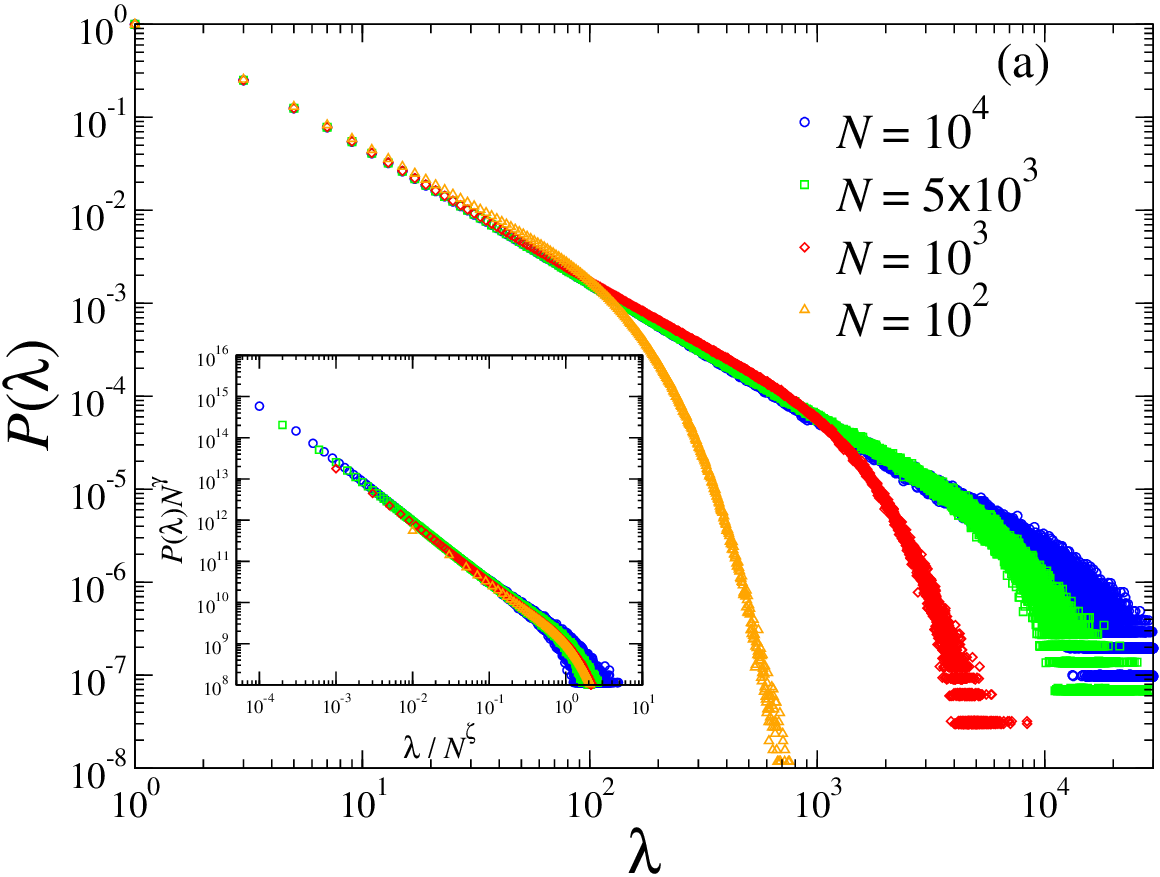}
\includegraphics[width=8cm]{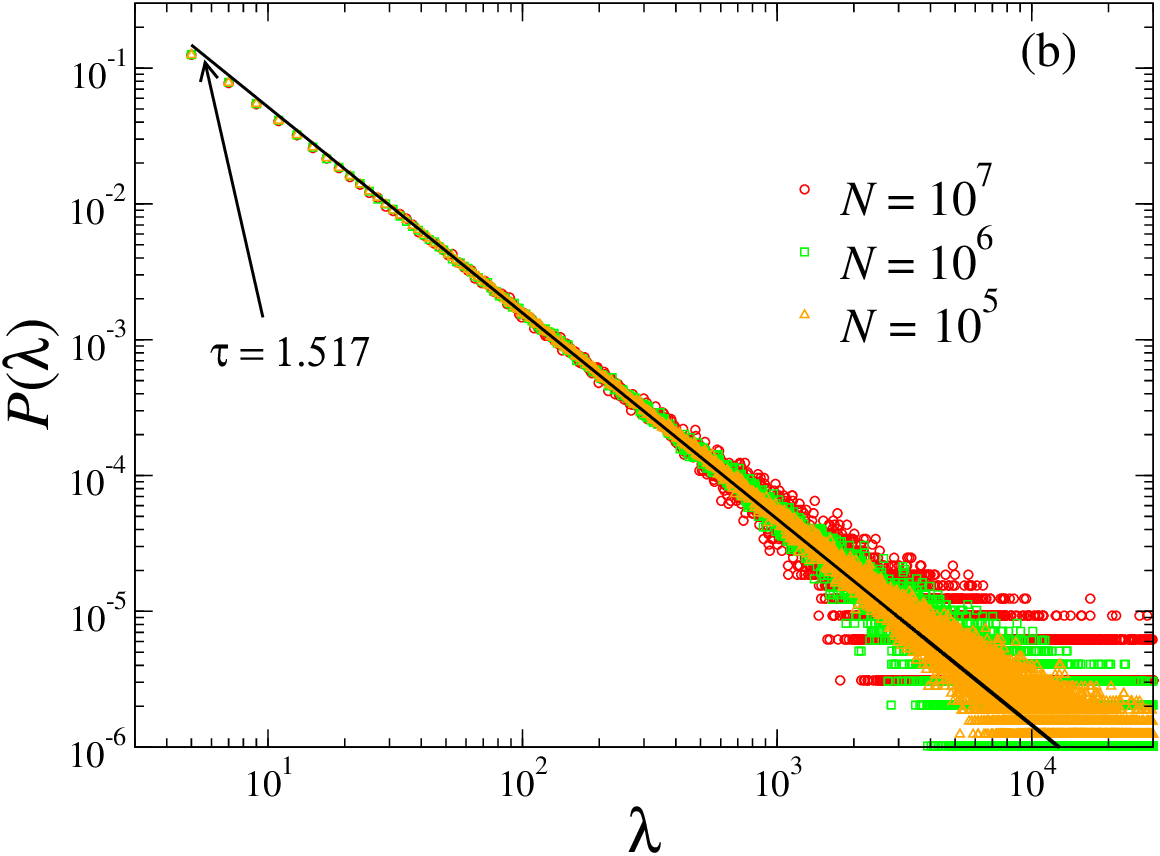}
\end{center}
\caption{(color online)  Fluctuation length distributions for the small $N$s and 
for the large $N$s groups are given in (a) and (b), respectively. 
In the inset of (a), we also present data collapse of finite size scaling given in Eq.~(\ref{eq:fss}) for small $N$s 
group. The critical exponents derived from the fit are $\gamma\simeq1.517$ and $\zeta=1$. The full black line in 
(b) represents the fitting curve of the distribution with slope $\tau\simeq1.517$ which has been obtained by 
maximum likelihood estimation. The distributions have an arbitrary normalization such that $P(\lambda=1)=1$.}
\label{fig:small-large}
\end{figure*}

%%%%%%%%%%%%%%%%%%%%%%%%%%%%%%%%%
%\section{\label{sec:PDF} Distribution of fluctuation length and returns }
%%%%%%%%%%%%%%%%%%%%%%%%%%%%%%%%%
{\it Distribution of fluctuation length and returns:}$\,\,$  As it was first demonstrated by BTW sandpile model, 
a generic signature of SOC is the presence of a power-law as well as finite size scaling in the size or the duration 
distribution of the avalanches. Recently, a power-law regime following an exponential tail in the fluctuation length 
distribution for the Ehrenfest's dog-flea model has been reported for a very limited system size 
(\emph{i.e.,} $N=2500$) \cite{Nagler-PRE}. In our paper, in order to analyze the SOC in the dog-flea model through 
the fluctuation length distribution we simulate the corresponding stochastic process for seven different values of $N$ 
namely, $N=10^{2},\,10^{3},\,5\times10^{3},\,10^{4},\,10^{5},\,10^{6},$ and $10^{7}$. For convenience, let us group 
the first four different system sizes as ``\emph{small $N$s}'' and the remaining sizes as ``\emph{large $N$s}''. 
In Fig.~\ref{fig:small-large}(a) and (b) we plot the distribution of the fluctuation length time-series $\lambda(t)$ 
for the small $N$s and large $N$s, respectively. In order to have good statistics $10^{9}$ fluctuations for the small 
$N$s group and $2\times10^{9}$ fluctuations for the large $N$s group have been considered. In both cases the 
fluctuation distributions have a power-law regime, $P(\lambda)\sim\lambda^{-\tau}$ while in the small $N$s group the 
power-law regime is followed by an exponential decay because of the finite-size effect. For the small $N$s group one 
can control if the fluctuation length distribution $P(\lambda)$ obeys the following finite size scaling behavior,
\begin{equation}\label{eq:fss}
P(\lambda)\sim\frac{1}{N^{\gamma}}f\left(\frac{\lambda}{N^{\zeta}}\right),
\end{equation}
where $f$ is a suitable scaling function and $\gamma$ and $\zeta$ are critical exponents describing the scaling of the 
distribution function. In the inset of Fig.~\ref{fig:small-large}(a), a clear data collapse of $P(\lambda)$ is shown 
for the small $N$s group (\emph{i.e.}, $N=10^{2},\,10^{3},\,5\times10^{3},\,$ and $10^{4}$). This data collapse 
indicates that the fluctuation length distributions of small $N$s satisfy the finite size scaling hypothesis very well. 
The obtained critical exponents are $\gamma\simeq1.517$ and $\zeta=1$. As it is seen from Fig.~\ref{fig:small-large}(b), 
these values of critical exponents are in agreement with the finite size scaling hypothesis since for asymptotically 
large $N$, $P(\lambda)\sim\lambda^{-\tau}$ with $\tau=\gamma/\zeta\simeq1.517$. The value of $\tau$ is obtained by the 
maximum likelihood estimation (MLE) and this method enables us to determine this exponent of the model as accurate as 
$\pm1.156\times10^{-5}$ \cite{Newman}.

%------------------------------ Figure 1 ----------------------------

%------------------------ Figure 2 --------------------------
\begin{figure*}[t]
\begin{center}
\includegraphics[width=8cm]{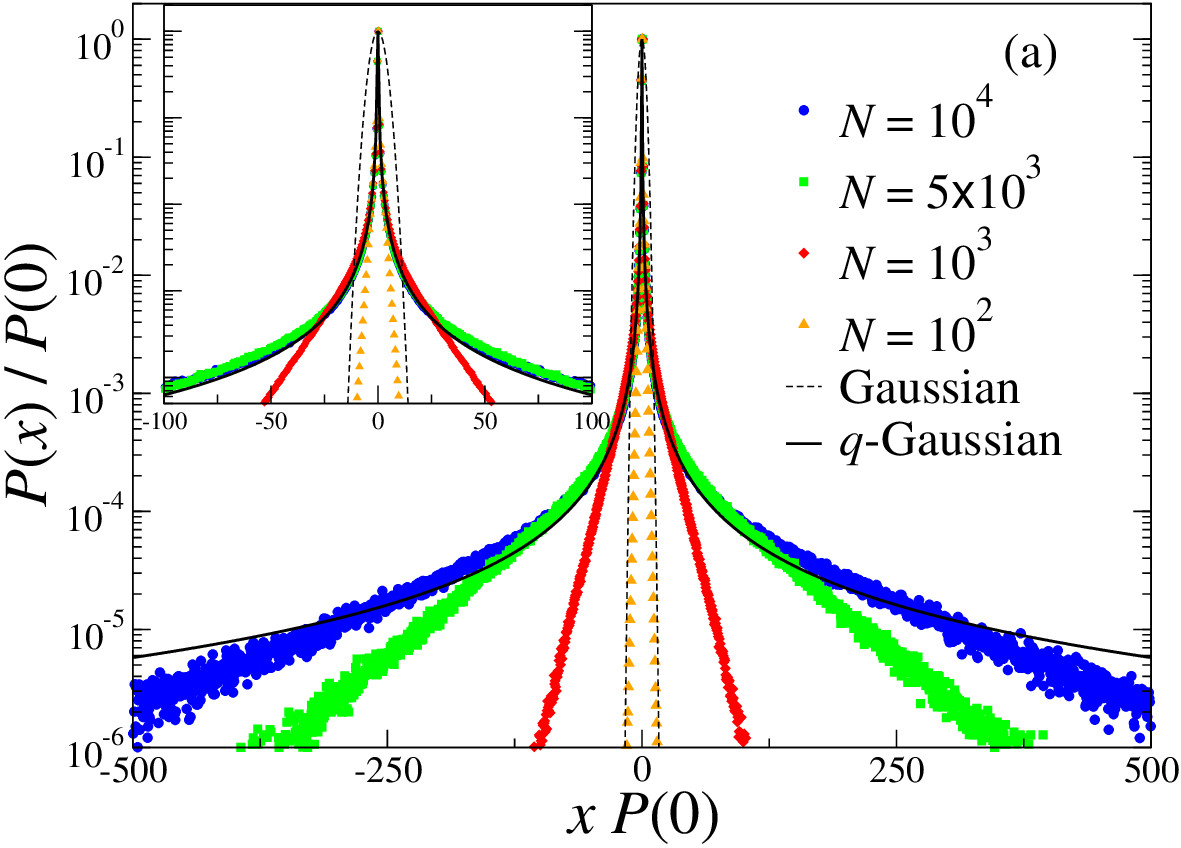}
\includegraphics[width=8cm]{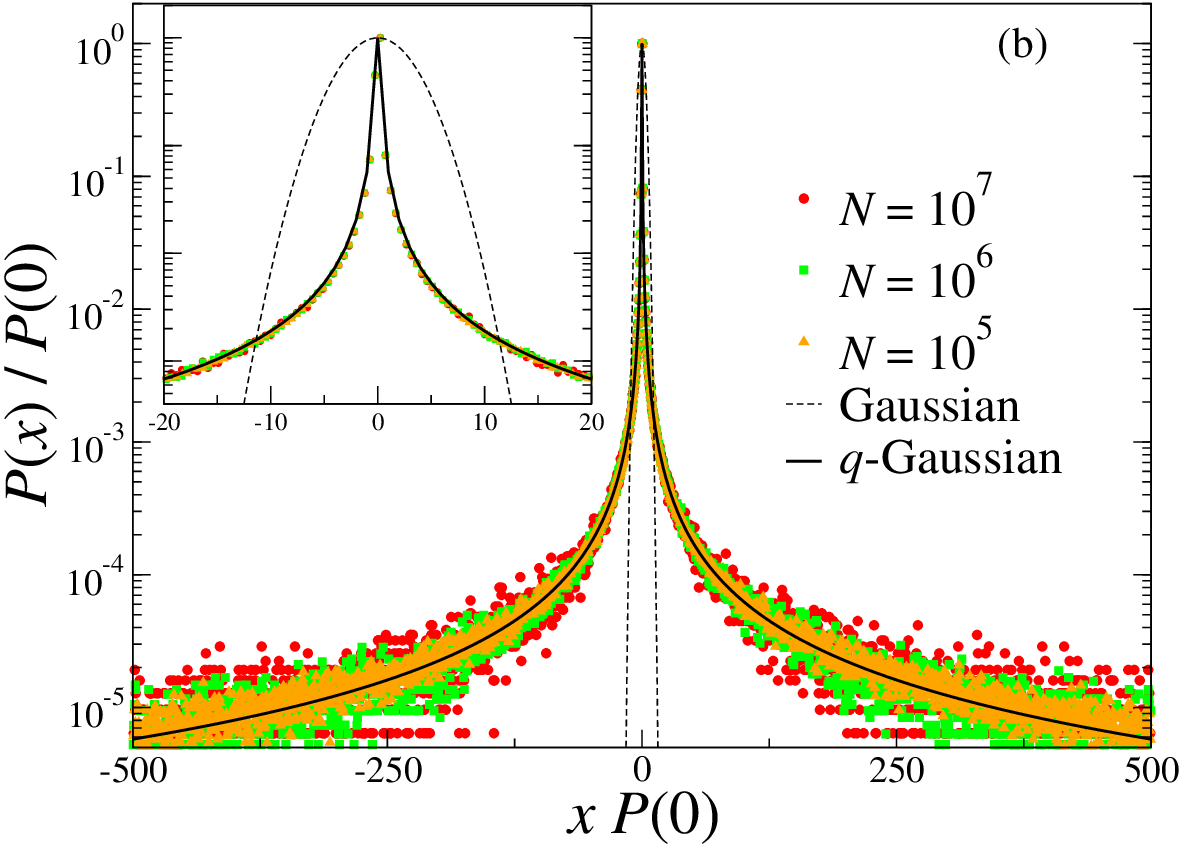}
\end{center}
\caption{(color online)  The distributions of returns, \emph{i.e.}, the fluctuation length differences 
$\Delta\lambda(t)=\lambda(t+1)-\lambda(t)$, normalized by introducing the variable 
$x=\Delta\lambda-\langle\Delta\lambda\rangle$ are shown in (a) for the small $N$s group and in (b) for the large $N$s 
group. For comparison, standard Gaussian and $q$-Gaussian curves are drawn by black dashed and full lines, respectively. 
See text for further details. In insets, the central parts of the distributions are emphasized. }
\label{fig:small-large2}
\end{figure*}
%-----------------------------------------------------------  

Now we are at the position to introduce the distribution of returns, i.e., the differences between fluctuation lengths 
obtained at consecutive time steps, as $\Delta\lambda(t)=\lambda(t+1)-\lambda(t)$. It should also be noted that, 
in order to have zero mean, the returns are normalized by introducing the variable $x$ as 
\begin{equation}\label{eq:x}
x=\Delta\lambda-\langle\Delta\lambda\rangle,
\end{equation}
where $\langle\cdots\rangle$ stays for the mean value of the given data set. The signal of the distribution of returns 
reveals very interesting results on the criticality of the dog-flea model. This approach is used in recent studies on 
turbulence \cite{Menech-Physica-Beck-PRE72} and the time-series of 
real earthquakes \cite{CarusoPRE75}.
%------------------------- Figure 3------------------------------------
\begin{figure} [t]
\begin{center}
\includegraphics[width=8cm]{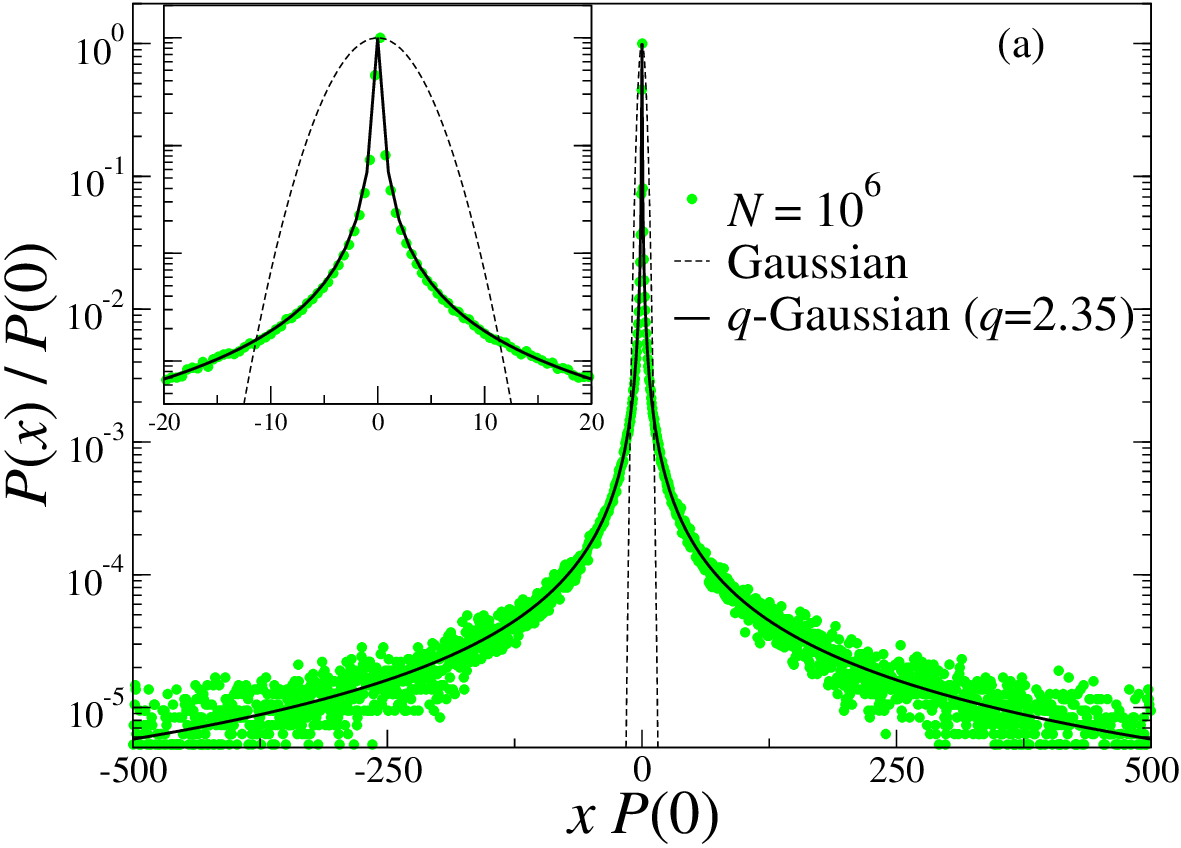}
\includegraphics[width=8cm]{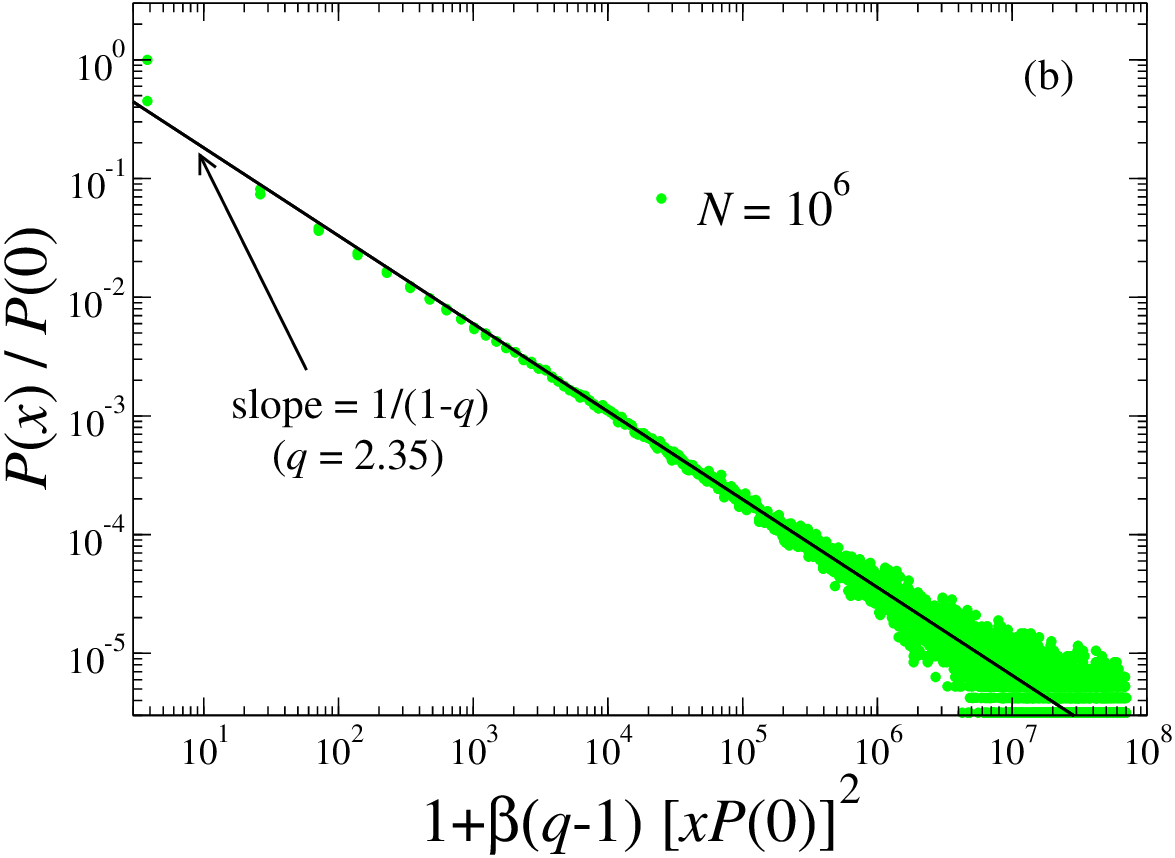}
\end{center}
\caption{(color online) (a) Distribution of returns for a representative case of large $N$s group ($N=10^6$) is given 
by full green circles. The $q$-Gaussian curve with $q=2.35$ and $\beta=35$ is shown by full black line. This value of 
the $q$ is obtained by substituting $\tau=1.517$ into Eq.~(\ref{eq:q-tau}). A standard Gaussian curve is drawn by 
dashed black line for comparison. In the inset, the central part of the distribution is given in order to emphasize 
that the distribution approaches almost perfectly to the $q$-Gaussian not only in the tails but also in the center. 
(b) In order to better visualize how well the used $q$-Gaussian approaches to the distribution, we plot the same 
$P(x)$ versus $1+\beta(q-1)x^{2}$. A straight line with a slope $1/(1-q)$ is expected for a perfectly $q$-Gaussian 
shaped distribution. Data points (green circles) and the slope with $q=2.35$ (black line) constitute a clear evidence 
towards this tendency.}
\label{fig:fitting}
\end{figure}
%---------------------------------------------    

In Fig.~\ref{fig:small-large2}, we plot the distribution of the returns $\Delta\lambda(t)$ obtained from $10^{9}$ 
fluctuations for each different system sizes in the small $N$s group (a), whereas in the group of large $N$s 
(b) $2\times10^{9}$ fluctuations are considered. What is common for both cases is that none of them has  return 
distributions which can be approached by a Gaussian. As the system size $N$ increases, leading to a longer power-law 
regime in the fluctuation length distribution, the return distribution curves become to exhibit a convergence to a kind 
of fat tailed distribution. When the system size is large enough, the exponential decay of the fluctuation length 
distribution (see Fig.~\ref{fig:small-large}(b)) is postponed to larger sizes and the finite size effects get invisible 
up to more than four decades. In this case the distribution of the returns can be fitted by a $q$-Gaussian given by 
\begin{equation}\label{eq:q-gauss}
P(x)=P(0)[(1+\bar{\beta}(q-1)x^{2}]^{1/(1-q)}, 
\end{equation}
where $\bar{\beta}$ characterizes the width of the distribution and $q$ is the index of nonextensive statistical 
mechanics \cite{Tsallis88-NonExEnt} (black full lines in Figs.~\ref{fig:small-large2}(a) and (b)). 
In Eq.~(\ref{eq:q-gauss}), $q\neq1$ indicates a departure from the Gaussian shape while normal Gaussian distribution 
can be recovered again in the $q\rightarrow 1$ limit.
Here, it is worth mentioning that our results in Fig.~\ref{fig:small-large2} clearly show the connection 
between criticality and the appearance of $q$-Gaussian, namely, wider the critical regime persists, longer the 
tails of returns distribution follow $q$-Gaussian. This kind of interpretation might also be useful in 
understanding the difference between two recent experimental works on velocity distributions in optical 
lattices \cite{optik1, optik2}. In \cite{optik1}, velocity distributions are found to approach a double-Gaussian 
shape, whereas in \cite{optik2} they are reported to converge to a $q$-Gaussian. The reason for this 
discrepancy seen in the results of essentially the same experiment might be that in the latter the system 
may be set exactly at the criticality, whereas in the former it is not.

At this point, we should recall the important result reported by Caruso \emph{et al.} \cite{CarusoPRE75} relating the $\tau$ exponent of the avalanche size distribution with the $q$ parameter of the $q$-Gaussian. As it was emphasized in their work, if there is no correlation between the size of two events, the probability of obtaining the difference $\Delta\lambda=\lambda(t+\delta)-\lambda(t)$ ($\delta$ is an integer describing the correlation length and in our case $\delta=1$) is given by
\begin{equation}\label{eq:hyper}
P(\Delta\lambda)=K\frac{\epsilon^{-(2\tau-1)}}{2\tau-1}{_2F_1}\left(\tau,2\tau-1;2\tau; -\frac{|\Delta\lambda|}{\epsilon}\right),
\end{equation}
where $K$ is a normalization factor, $\epsilon$ is a small positive value and ${_2F_1}$ is the hypergeometric function. The curve of this $\tau$ dependent probability density function $P(\Delta\lambda)$ can be approached by means of $q$-Gaussian with $\epsilon$-independent $q$ value. In Ref.~\cite{CarusoPRE75}, by evaluating Eq.~(\ref{eq:hyper}) for various values of $\tau$, a relation between the power-law exponent $\tau$ and $q$ is reported as
\begin{equation}\label{eq:q-tau}
q=e^{1.19\tau^{-0.795}}.
\end{equation}       
Although this relation is obtained in \cite{CarusoPRE75} by Caruso \emph{et al.}, they could not check its validity since the earthquake data that they analyzed was not adequate to obtain the $\tau$ value with high precision. Consequently, they still used $q$ parameter as a fitting parameter. On the other hand, since the power-law exponent is very accurate in our case, we can substitute its value ($\tau=1.517$) obtained by MLE into Eq.~(\ref{eq:q-tau}) which gives the $q$ value as $q=2.35$. This value is obviously the one that we should use in the $q$-Gaussian to check whether the return distribution can be approached by this. It is worth mentioning here that the $q$ parameter is not a fitting parameter anymore. In Fig.~\ref{fig:small-large2} we also include this result together with a Gaussian curve for comparison. It is clear that, for very small $N$s, the convergence to $q$-Gaussian is only in the central part 
(see the inset of Fig.~\ref{fig:small-large2}(a)), whereas it develops more and more towards the tails as $N$ increases. Eventually, for large enough $N$s for which finite size effects are invisible inside the obtained region, the $q$-Gaussian curve is perfectly approached including the center and tails. 

In order to further strengthen our results, we consider one of the appropriate system size ($N=10^{6}$) separately in Fig.~\ref{fig:fitting}. A very clear convergence of the return distribution to the $q$-Gaussian can be seen \emph{everywhere} for the available data (including the very central part, see the inset of Fig.~\ref{fig:fitting}(a)). Moreover, to check how well the obtained $q$-Gaussian curve approaches the returns distribution, a log-log plot of Eq.~(\ref{eq:q-gauss}) is given in Fig.~\ref{fig:fitting}(b). A perfect straight line with the slope $1/(1-q)$ is the expected behavior for this type of representation if the curve is an exact $q$-Gaussian and as it is seen very clearly, the behavior of the return distribution fulfills this tendency exhibiting a seven decade power-law with the slope $1/(1-q)$ 
which gives the already obtained $q$ value, $q=2.35$. 

%%%%%%%%%%%%%%%%%%%%%%%%%%%%%%%%%%%%%%%%%%
%\section{\label{sec:Conc} Conclusion}
%%%%%%%%%%%%%%%%%%%%%%%%%%%%%%%%%%%%%%%%%%
{\it Conclusion:}$\,\,$  We analyze the SOC in the Ehrenfest's dog-flea model through the probability 
distributions of the fluctuation length (avalanche size distributions) and of the differences between 
the fluctuation lengths at subsequent time steps (returns distributions) by simulating the stochastic 
process of the model. Our extensive simulations enable us to determine the power-law exponent $\tau$ 
of the avalanche size distribution with an extreme precision. 
Then, the behavior of the returns distributions is analyzed and numerically shown that it converges 
to a $q$-Gaussian with $q=2.35$, a value coming directly (and \emph{a priori}) from Eq.~(\ref{eq:q-tau}) 
which makes $q$ parameter to be related to one of the well known power-law exponents of such model systems 
(which means that $q$ is not a fitting parameter anymore). 
This is the main result of the present letter and important from (at least) three 
point of view: 
(i)~this constitutes the first reliable verification of Caruso \emph{et al.} relation since, 
due to insufficient data set of earthquakes, they were unable to provide a clear evidence for 
their own relation; 
(ii)~this result is achieved using a simple, prototype SOC model (different from 
the one used by Caruso \emph{et al.}) which can be considered as the first clue on the generality of 
these results rather than being specific only to this model;  
(iii)~this treatment makes the $q$ parameter of the $q$-Gaussian to be determined {\it a priori} 
which constitutes a rather rare achievement in the literature due to technical difficulties. 
>From the analysis of return distributions from small $N$s to large $N$s, it is shown that the convergence 
to appropriate $q$-Gaussian starts from the central part and gradually develops towards the tails as $N$ 
increases. This is a kind of expected behavior since, from our simulations it is also evident that the 
power-law regimes of the avalanche size distributions for small $N$s are followed by exponential decays 
due to finite size effects and this obviously deteriorates the true behavior. Of course, for large enough 
$N$s, this effect is postponed further and further to avalanche sizes that are not inside the region we are 
considering. Moreover, one could conclude that, as $N\rightarrow\infty$ the power-law regime of avalanche 
size distribution is expected to continue forever, then the corresponding return distribution appears to 
converge to the $q$-Gaussian for the \emph{entire} region.
Finally, it is worth to mention that the behavior observed and reported here for the zero dimensional 
prototype SOC model of Ehrenfest is by no means specific and limited to this model, but seems to appear 
as a rather common phenomenon for several SOC models \cite{bb-ut-inprep}. \\

%%%%%%%%%%%%%%%%%%%%%%%%%%
%\acknowledgements 
%%%%%%%%%%%%%%%%%%%%%%%%%%
This work has been supported by TUBITAK (Turkish Agency) under the Research Project number 104T148.

\bibliography{FleasPRL}

\end{document}